# Brand effect versus competitiveness in hypernetworks

Jin-Li Guo[1,2,*] Qi Suo[1]

[1]*Business School, University of Shanghai for Science and Technology, Shanghai, 200093, China*

[2]*Center for Supernetwork Research, University of Shanghai for Science and Technology, Shanghai, 200093, China*

A few of evolving models in hypernetworks have been proposed based on uniform growth. In order to better depict the growth mechanism and competitive aspect of real hypernetworks, we propose a model in term of the non-uniform growth. Besides hyperdegrees, the other two important factors are introduced to underlie preferential attachment. One dimension is the brand effect and the other is the competitiveness. Our model can accurately describe the evolution of real hypernetworks. The paper analyzes the model and calculates the stationary average hyperdegree distribution of the hypernetwork by using Poisson process theory and a continuous technique. We also address the limit in which this model has a condensation. The theoretical analyses agree with numerical simulations. Our model is universal, in that the standard preferential attachment, the fitness model in complex networks and scale-free model in hypernetworks can all be seen as degenerate cases of the model.

**Keywords:** Hypergraph; hypernetwork; complex network; scale-free network; power-law distribution.

**The hypernetworks based on hypergraphs can depict complex relationships between the objects in real systems. However, there are little literatures on them. In this paper we propose and analyze a hypernetwork model. The growth mechanism is non-uniform. Namely, the number of nodes of each hyperedge obeys a certain distribution rather than a fixed value. Three dimensions, including hyperdegrees, fitness and competitiveness underlie preferential attachment. The stationary average hyperdegree distribution of the hypernetwork shows that this competition for hyperedges translates into multiscaling. This can help us understand the evolution of many competitive systems in nature and society.**

## I. INTRODUCTION

An upsurge of research into the complex networks has swept across the academia since the Watts-Strogatz model [1] and Barabási–Albert model [2] were proposed at the end of 20th century. During the past decade, complex network theory, as a useful representation of natural and social systems, has received increasing interests from researchers (e.g. physics, biology, computer, economics and sociology). Scholars have studied on the topological properties of complex networks. In addition, they have also proposed a number of models [3, 4]. Generally speaking, in complex networks, the nodes represent different individuals and the edges represent the relationships between the nodes, and each edge could only associates with two nodes. In real networks, there are large numbers of edges and nodes, the diversification of the edge types and the complexity of the network structure. Therefore, the complex networks could no longer fully depict the characteristics of complex systems [5]. For example, in the scientific collaboration network, an edge can only describe collaboration between two authors, but we do not know whether two or more authors were coauthors of the same paper or not. Besides, Lv and Medo *et al.* [6] discussed the system containing three kinds of nodes (i.e. users, commodities and tag recommendation), which can hardly be depicted by a bipartite graph. To resolve this issue, the hypergraph can be used to give an exact representation of the full structure of a collaborative tagging system. Berge [7] proposed the basic concepts and properties of the hypergraph theory, with a hyperedge containing arbitrary nodes. The hypernetwork, based on the hypergraph theory, would effectively reveal the influence and the interaction of a variety of nodes. For example, in the scientific collaboration hypernetworks, authors and papers are regarded as nodes and hyperedges, respectively. Another example, a chemical reaction can be viewed as a hyperedge while nodes are chemicals. Similarly, in ecological hypernetworks, nodes represent species and hyperedges represent groups of species that compete for common prey. The competitive hypernetworks reflect the state of the competition between species. Finally, in supply chain hypernetworks, nodes represent the suppliers, manufacturers and consumers, and hyperedges could be seen as the trade activities among different types of entities. Therefore, the hypernetwork provides such a powerful approach for accurately depicting the real-life network that it will cause future interests of researches.

---

* To whom correspondence should be addressed. E-mail: phd5816@163.com

Recently, some scholars have studied the topology properties of hypernetworks. Estrada *et al.* [8] studied the subgraph centrality and the clustering coefficient in hypernetworks. Ghoshal *et al.* [9] studied random hypergraphs and their applications. Zlatic et al. [10] extended tripartite hypergraph model by defining additional quantities such as edge distributions, vertex similarity and correlations as well as clustering. Ma *et al.* [11] constructed the hypernetwork model of internet public opinion and put forward some indexes such as node superdegree, superedge-superedge distance and superedge overlap. Quan *et al.* [12] discussed the node importance in hypernetwork model on the basis of the node hyperdegree and betweenness. Currently, scholars have studied some important topological properties in hypernetworks.

However, the literatures on the evolving hypernetworks are limited. Although Ni *et al.* [13] analyzed topological properties of the evolving hypernetwork of Wiki ontology, the model in Ref. [13] is a special complex network model in Ref. [14] rather than a hypernetwork model. Zhang *et al.* [15] established a dual preferential attachment mechanism growth model, based on the users' background knowledge, objects and labels. Pei *et al.* [16] studied a dynamic model with triangular structure, obtaining the theoretical solution by mean-field theory. Wang *et al.* [17] proposed an evolving model based on the growth and preferential attachment mechanisms. That is, at every time step, $m$ new nodes are added, and a selected old node and these $m$ nodes construct a new hyperedge. Hu *et al.* [18] proposed another kind of evolving model that at each time step they added a new node, a new hyperedge encircling the new node and $m$ existing nodes. Wu *et al.* [19] discussed the evolving model based upon the same mechanisms as in Ref. [18]. We proposed a scale-free hypernetwork [20] unifying the models in Refs. [17-19].

From the above literature, the common characteristics of hypernetwork models are as follows. (1) The growth mechanism: at every time step a number of new nodes are added and encircled with the existing nodes by a new hyperedge. The number of nodes in each hyperedge is fixed, namely, the hypernetwork is uniform. (2) The preferential attachment mechanism: the preferential probability is proportional to the hyperdegree of the selected node. The higher the node hyperdegree is, the more probability it will be connected. This preferential attachment mechanism means that the hyperdegree of the old nodes will be higher than that of the new nodes, namely the phenomenon that "the rich get richer".

However, the mechanisms above could not perfectly depict the evolving process of the real-life network. There are two problems as follows. (1) It is clearly that the uniform hypernetworks may not better describe the realistic systems. (2) Although nodes with higher hyperdegree will have more probability to be connected, the existing hypernetwork models neglect other dimensions. The hyperdegree of the node is regarded as 'the brand effect'. The principle that 'the brand effect is attractive' underlies preferential attachment, which is a common explanation for the emergence of scaling in hypernetworks. However, the probability of obtaining hyperedge not only depends on the node's brand effect, but also closely relates with its own fitness and competitiveness. For example, some young web pages attract a large number of links in a short time, and even obtain more links than the old ones for its splendid contents and good publicity in World Wide Web. On account of revealing important scientific discoveries, some later-published scientific papers even get more citations than earlier papers in a short-term. Therefore, each node is assigned with an adaptive capacity parameter (fitness) [21] and competitiveness. These parameters can reflect its ability of gaining a new hyperedge. Thus it can be used to depict the property of hypernetworks. Attentions to the competitive hypernetworks would certainly motivate the studies of new models in network science. Whether there is a hypernetwork model can depict the non-uniform evolving mechanism combining with brand effect, fitness and competitiveness. The purpose of this paper is to answer this question.

The rest of this paper is organized as follows. The next section introduces the concepts of the hypergraph and the hypernetwork. In section 3, we introduce the scale-free hypernetwork model and develop a co-evolution model with the brand effect and competitiveness. We obtain stationary average hyperdegree distribution by using Poisson process theory and a continuous technique. In section 4, the numerical simulations show that our analysis method is feasible and the result is consistent with theoretical forecast. Finally, the summary is given in section 5.

## II. THE CONCEPT OF HYPERNETWORK

Denning [23] proposed the concept of 'Supernetworks': they are organized by networks. Nagurney *et al.* [24] further elaborated it upon that supernetworks are 'networks of networks' which have large scale, complex connections and nested networks. When dealing with problems including financial, informational, and logistical flows, Nagurney *et al.* presented supernetworks to depict networks which 'above and beyond the existing network'. 'Above and beyond' means that networks nest networks and consist of virtual edges, nodes and flows [5].

Another concept is the hypernetwork based on hypergraphs. It was proposed by Berge in 1970 [7]. In a graph a link relates only two nodes, but the edges of the hypergraph, known as hyperedges, can relate more than two nodes. Estrada *et al.* identified the hypernetwork could be described by hypergraphs [23]. The mathematical definition of the hypergraph is as follows. Let $V = \{v_1, v_2, \cdots, v_n\}$ be a finite set, and let $E_i = \{v_{i_1}, v_{i_2}, \cdots, v_{i_k}\}$ $(v_{i_j} \in V, j = 1, 2, \cdots, k)$, $E^h = \{E_1, E_2, \cdots, E_m\}$ be a family of subsets of $V$. The pair $H = (V, E^h)$ is called a hypergraph. The elements in $V$ are called the nodes, and $E_i$ $(1, 2, \cdots, m)$ is a set of non-empty subsets of $V$ called a hyperedge. In a hypergraph, two nodes are said to be adjacent if there is a hyperedge that contains both of these nodes. Two hyperedges are said to be adjacent if their intersection is not empty. If $|V|$ and $|E^h|$ are finite, $H$ is a finite hypergraph.

If $|E_i| = u$ $(i = 1, 2, \cdots, m)$, $H = (V, E^h)$ is an $u$-uniform hypergraph. If $|E_i| = 2$, $(i = 1, 2, \cdots, m)$, $H = (V, E^h)$ degrades to a graph.

Based on the above definitions, we can give mathematical definition of the hypernetwork. Suppose $\Omega = \{(V, E^h) | (V, E^h) \text{ is a finite hypergraph}\}$ and $G$ is a map from $T = [0, +\infty)$ into $\Omega$; for any given $t \geq 0$, $G(t) = (V(t), E^h(t))$ is a finite hypergraph. The index $t$ is often interpreted as time. A hypernetwork $\{G(t), t \in T\}$ is a collection of hypergraphs. The hyperdegree of $v_i$ is defined as the number of hyperedges that connect to node $v_i$.

Suppose that $N(t) = |V(t)|$, $M(t) = |E^h(t)|$, $\lim_{t \to \infty} E[N(t)] = N$ (finite or infinite), where $E[N(t)]$ denotes the average number of nodes in the hypernetwork at the time $t$. For any given finite hypergraph $(V, E^h)$ and $t \geq 0$, we take $G(t) = (V, E^h)$. Thus, the hypernetwork is a generalization of the hypergraph.

## III. MODEL DESCRIPTION AND ANALYSIS

### 3.1 The scale-free hypernetwork evolving model

The scale-free hypernetwork evolving model is as follows. (1) The hypernetwork starts with $m_0$ nodes and a hyperedge including all these $m_0$ nodes. Suppose that nodes arrive at the system in accordance with a Poisson process having rate $\lambda$. If a new batch of $m_1$ nodes is added to the network at time $t$, a new hyperedge is formed by connecting this batch of $m_1$ nodes and $m_2$ previously existing nodes, totally $m$ new hyperedges are constructed with no repetitive hyperedges ($mm_2 \leq m_0$). (2) At time $t$, the probability $W$ that a new node will connect to the $j$th node of the $i$th batch, is proportional to the hyperdegree $h_j(t, t_i)$ of that node, such that

$$W(h_j(t, t_i)) = \frac{h_j(t, t_i)}{\sum_{ij} h_j(t, t_i)} \quad , \tag{1}$$

where $t_i$ denotes the time at which the $i$th batch of nodes is added to the network, that is to say, the birth time of the $i$th batch of nodes is $t_i$.

The stationary average hyperdegree distribution of the hypernetwork evolving model [20] is $P(k) \approx \frac{1}{m}(\frac{m_1}{m_2} + 1)(\frac{m}{k})^{\frac{m_1}{m_2}+2}$, and the hyperdegree distribution exponent is $\gamma = \frac{m_1}{m_2} + 2$.

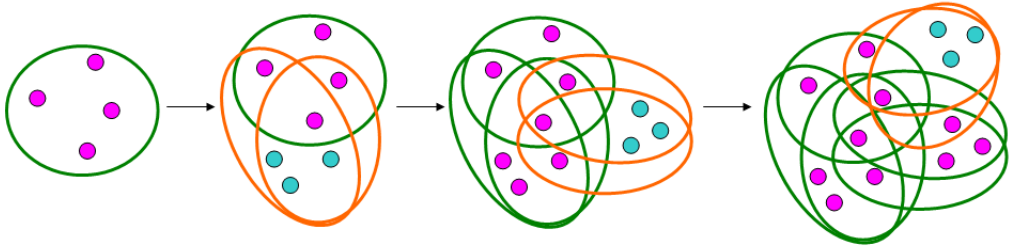

FIG. 1.　Schematic description of the evolving process of an uniform hypernetwork model

**3.2 The hypernetwork model with the brand effect and competitiveness**

The hypernetwork model with the brand effect and competitiveness satisfies the following two conditions. (1) The hypernetwork starts with $m_0$ nodes and a hyperedge including all these $m_0$ nodes. Suppose that nodes arrive at the system in accordance with Poisson process $N(t)$ having rate $\lambda$. Each batch of nodes entering the network is tagged with its own competitiveness $\xi_i$ and fitness $y_i$, where $\xi_i$, $y_i$ are taken from given distributions $G(x)$ and $F(y)$, respectively. Here $c = \int x dG(x)$ and $E[y] = \int x dF(x)$ are finite. If a new batch of $m_1$ nodes is added to the network at time $t$, $\eta_{N(t)}$ is the positive integer that is sampled from the total which has a distribution $Q(n)$ and $m_1 = \sum_n n Q(n)$ is finite. A new hyperedge is formed by connecting this batch of $\eta_{N(t)}$ nodes and $m_2$ previously existing nodes, and totally $m$ new hyperedges are constructed with no repetitive hyperedges ($mm_2 \leq m_0$); (2) At time $t$, the probability that a new node will connect to the $j$th node of the $i$th batch, is proportional to the hyperdegree $h_j(t,t_i)$, competitiveness $\xi_i$ and fitness $y_i$ of that node, such that

$$W(h_j(t,t_i)) = \frac{y_i h_j(t,t_i) + \xi_i}{\sum_{ij}(y_i h_j(t,t_i) + \xi_i)}, \qquad (2)$$

Where $t_i$ denotes the time at which the $i$th batch of nodes is added to the network.

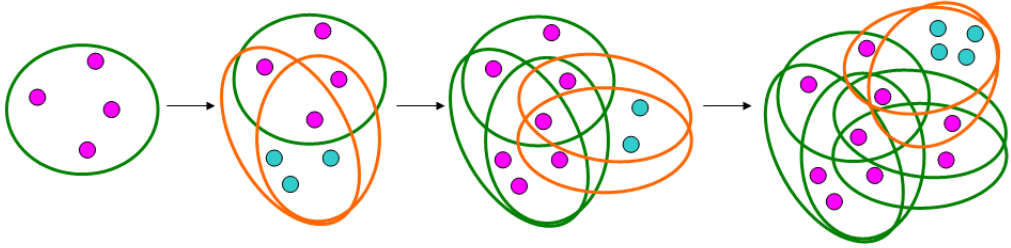

FIG. 2.  Schematic description of the evolving process of a non-uniform hypernetwork model

Suppose $h_j(t,t_i)$ is a continuous real variable, the rate at which $h_j(t,t_i)$ changes is expected to be proportional to probability $W$. Consequently, $h_j(t,t_i)$ satisfies the dynamic equation

$$\frac{\partial h_j(t,t_i)}{\partial t} = mm_2 \lambda \frac{y_i h_j(t,t_i) + \xi_i}{\sum_i (y_i h_j(t,t_i) + \xi_i)}. \qquad (3)$$

Let

$$B = \lim_{t \to \infty} \frac{1}{\lambda t} \sum_{ij} y_i h_j(t,t_i) + c. \qquad (4)$$

$N(t)$ denotes the total number of batches of nodes by time $t$. According to the Poisson process theory [3], $E[N(t)] \approx \lambda t$. Thus, for sufficiently large $t$, we have

$$B \approx \frac{1}{\lambda t} \sum_{ij}(y_i h_j(t,t_i) + \xi_i). \qquad (5)$$

Let $A = \frac{B}{mm_2}$, substituting Eq. (5) into Eq. (3), leading to

$$\frac{\partial h_j(t,t_i)}{\partial t} = \frac{y_i h_j(t,t_i) + \xi_i}{At}. \tag{6}$$

Since $h_j(t_i,t_i) = m$, solving Eq. (6), we have

$$h_j(t,t_i,y,\xi) = (m + \frac{\xi_i}{y_i})(\frac{t}{t_i})^{\beta(y_i)} - \frac{\xi_i}{y_i}, \tag{7}$$

Where $\beta(y) = \frac{y}{A}$

From Eq. (7), leading to

$$m_1 \int dF(y) \int dG(\xi) \int_0^t [(m + \frac{\xi}{y})(\frac{t}{s})^{\beta(y)} - \frac{\xi}{y}]\lambda ds = \sum_{i=0}^{N(t)} m(\eta_i + m_2) = m(m_1 + m_2)\lambda t$$

$A$ is determined by the following integral equation

$$mm_1 \int \frac{x}{x-y} dF(y) + cm_1 \int \frac{1}{x-y} dF(y) - m(m_1 + m_2) = 0, \tag{8}$$

and Eq. (8) is called the characteristic equation of hyperdegrees of the hypernetwork with the brand effect and competitiveness.

From Eq. (7), we obtain

$$P\{h_j(t,t_i,y,\xi) \geq k\} = P\{t_i \leq (\frac{y_i m + \xi_i}{y_i k + \xi_i})^{\frac{A}{y_i}} t\}, \quad t \gg t_i$$

Birth time $t_i$ of the $i$th batch of nodes is a random variable, according to the Poisson process theory, and it follows that $t_i$ will be the gamma random variable having parameters ($i$, $\lambda$), thus

$$P\{h_j(t,t_i,y,\xi) < k\} = e^{-\lambda t (\frac{y_i m + \xi_i}{y_i k + \xi_i})^{\frac{A}{y_i}}} \sum_{l=0}^{i-1} \frac{1}{l!} (\lambda t (\frac{y_i m + \xi_i}{y_i k + \xi_i})^{\frac{A}{y_i}})^l \tag{9}$$

From Eq. (9), we have

$$P\{h_j(t,t_i,y,\xi) = k\} \approx \frac{\lambda t A}{y_i m + \xi_i} (\frac{y_i m + \xi_i}{y_i k + \xi_i})^{\frac{A}{y_i}+1} e^{-\lambda t (\frac{y_i m + \xi_i}{y_i k + \xi_i})^{\frac{A}{y_i}}} \frac{1}{(i-1)!} \left[\lambda t (\frac{y_i m + \xi_i}{y_i k + \xi_i})^{\frac{A}{y_i}}\right]^{i-1} \tag{10}$$

From Eq. (10), we obtain the stationary average hyperdegree distribution as follows:

$$P(k) \approx A \int dF(y) \int \frac{1}{my + \xi} \left(\frac{my + \xi}{ky + \xi}\right)^{\frac{A}{y}+1} dG(\xi), \tag{11}$$

where $A$ is a solution of characteristic Eq.(8).

When $y_i = 1$, $\xi_i = c$, the hypernetwork model degenerates to an initial attractiveness model of hypernetworks, and the initial attractiveness model of complex networks [21] is its special case. From Eq.(8), we have

$$x = \frac{m_1}{m_2} + \frac{cm_1}{mm_2} + 1 \tag{12}$$

From Eq. (11), we obtain the stationary average hyperdegree distribution of an initial attractiveness model as follows:

$$P(k) \approx (\frac{m_1}{m_2} + \frac{cm_1}{mm_2} + 1)\frac{1}{m+c}\left(\frac{m+c}{k+c}\right)^{\frac{m_1}{m_2} + \frac{cm_1}{mm_2} + 2} \tag{13}$$

Eq. (13) shows that the initial attractiveness hypernetwork has the scale-free characteristic and its hyperdegree exponent is $\gamma = \frac{m_1}{m_2} + \frac{cm_1}{mm_2} + 2$.

When $\xi_i = 0$, Eq. (2) shows that the hypernetwork model degenerates to the fitness model of hypernetworks, and the fitness model of complex networks [22] is its special case. From Eq. (11) and characteristic Eq. (8), we conclude

$$P(k) \approx \frac{C}{m} \int \frac{1}{y} \left(\frac{m}{k}\right)^{\frac{C}{y}+1} dF(y) \qquad (14)$$

Where $C$ is a solution of the following integral equation:

$$m_1 \int \frac{x}{x-y} dF(y) = m_1 + m_2 \qquad (15)$$

When $y_i = 1$, $\xi_i = 0$, the hypernetwork model degenerates to the scale-free hypernetwork model in Ref. [20], and the hypetdegree exponet is $\gamma = \frac{m_1}{m_2} + 2$.

Therefore, our model unifies the fitness model, the competitiveness model and the scale-free hypernetwork model.

When $y_i$ and $\xi_i$ are taken from the uniform distributions over [0, 1], respectively, the stationary average hyperdegree distribution is as follows:

$$P(k) \approx A \int_0^1 dy \int_0^1 \frac{1}{my + \xi} \left(\frac{my + \xi}{ky + \xi}\right)^{\frac{A}{y}+1} d\xi, \qquad (16)$$

where $A$ is a solution of the following characteristic equation:

$$(mm_1 x + \frac{1}{2} m_1) \ln \left|\frac{x}{x-1}\right| - m(m_1 + m_2) = 0 \qquad (17)$$

When $y_i$ is taken from the uniform distributions over [0, 1] and $\xi_i = 1$, the stationary average hyperdegree distribution is as follows:

$$P(k) \approx A \int_0^1 \frac{1}{my+1} \left(\frac{my+1}{ky+1}\right)^{\frac{A}{y}+1} dy, \qquad (18)$$

where $A$ is a solution of the following characteristic equation:

$$(mm_1 x + m_1) \ln \left|\frac{x}{x-1}\right| - m(m_1 + m_2) = 0 \qquad (19)$$

Next we address the limit in which this model has a condensation, similarly to what happens for the fitness networks [25], and for growing weighted networks [26]. We get the hyperdegree distribution under the condition that Eq. (8) has a positive solution. Bose-Einstein condensation appears when Eq. (8) has no positive solution. When $\xi_i = 0 (i = 1,2,3,\cdots)$, Eq. (8) is equivalent to

$$\int_0^{+\infty} \frac{y}{x-y} dF(y) = \frac{m_2}{m_1}. \qquad (20)$$

We assign an energy $\varepsilon_i$ to each node, determined by its fitness $y_i$ through the relation

$$\varepsilon_i = -\frac{1}{\beta} \ln y_i, \qquad (21)$$

where $\beta$ is a parameter playing the role of inverse temperature. Let $\rho(y) = F'(y)$, then the probability density function of $\varepsilon_i$ is $\varphi(\varepsilon) = \beta \rho(e^{-\beta\varepsilon}) e^{-\beta\varepsilon}$. Substituting $y = e^{-\beta\varepsilon}$ into Eq. (20), leading to

$$\int_{-\infty}^{+\infty} \frac{e^{-\beta\varepsilon}}{x - e^{-\beta\varepsilon}} \varphi(\varepsilon) d\varepsilon = \frac{m_2}{m_1}. \qquad (22)$$

Since $x = A = \frac{B}{mm_2}$ is positive, we introduce the chemical potential $\mu$ as $x = e^{-\beta\mu}$, which allows us to

$$I(\beta,\mu) = \int_{-\infty}^{+\infty} \frac{1}{e^{\beta(\varepsilon-\mu)}-1} \varphi(\varepsilon)d\varepsilon = \frac{m_2}{m_1}. \qquad (23)$$

However, $I(\beta,\mu)$ defined in (23) takes its maximum at $\mu=0$, thus if $I(\beta,0) < m_2/m_1$ for a given $\beta$ and $\varphi(\varepsilon)$, Eq. (23) has no solution. The absence of a solution indicates that almost all nodes have a few of edges, connecting them to some "gel" nodes that have the rest of the edges of the network. It seems to be a well-known signature of Bose-Einstein condensation [25].

## IV. SIMULATIONS AND ANALYSIS

In the following simulations, the parameters are set as follows: the number of initial nodes $m_0 = 20$, the number of hyperedges $m=2$, the number of selected existing nodes $m_2 = 2$, the competitiveness $\xi_i = 1$, the fitness is taken from the uniform distributions over [0,1]. Numerical simulations are performed with hypernetwork scale $N = 300000$. Figure 3 to Figure 5 shows the case of the same parameters described above and different values of $\eta_{N(t)}$ (the number of new nodes), which is random selected from 1~3, 1~5 and 1~11, respectively. The simulation results are showed in double-logarithmic axis. The points plotted are the hyperdegree distribution of the simulations, and straight line indicates theoretical prediction form Eq. (18). By the evolution mechanism of the model, we know that the number of links, its own competitiveness $\xi_i$ and fitness $y_i$ jointly determine the attractiveness and evolution of a node. Thus the ability for a node to acquire new hyperedges is not equal. As new nodes appear, they tend to connect to the most attractive nodes, and these ones thus acquire more links over time than their less attractive neighbors. And this process will generally favor the most attractive nodes which results in that a tiny fraction of the most attractive nodes will acquire a certain number of links. As the figures show, all hyperdegree distributions exhibit a power-law form as $\eta_{N(t)}$ changes. And the theoretical prediction of the hyperdegree distribution is in good agreement with the simulation results.

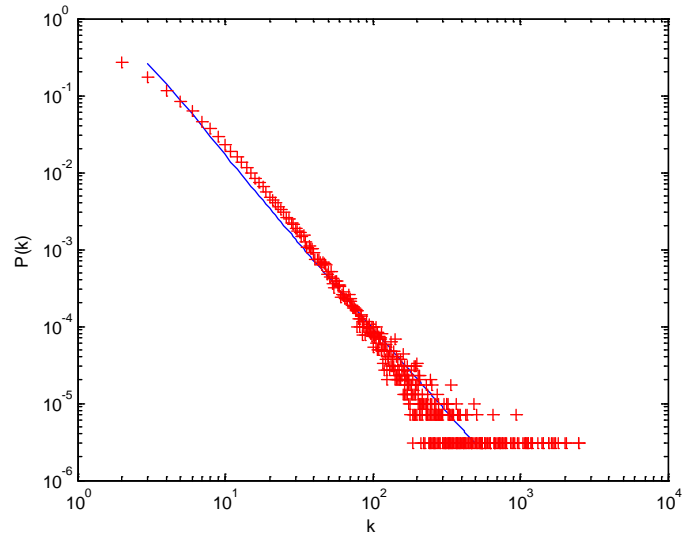

FIG. 3. The hypernetwork model ($\eta_{N(t)}$ is random selected from 1~3) simulation. + denotes the simulation result, the line denotes theoretical prediction.

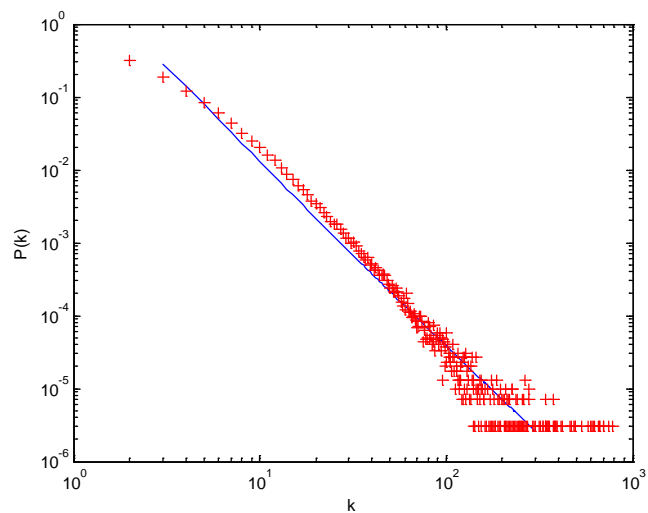

FIG. 4. The hypernetwork model ($\eta_{N(t)}$ is random selected from 1~5) simulation. +denotes the simulation result, the line denotes theoretical prediction.

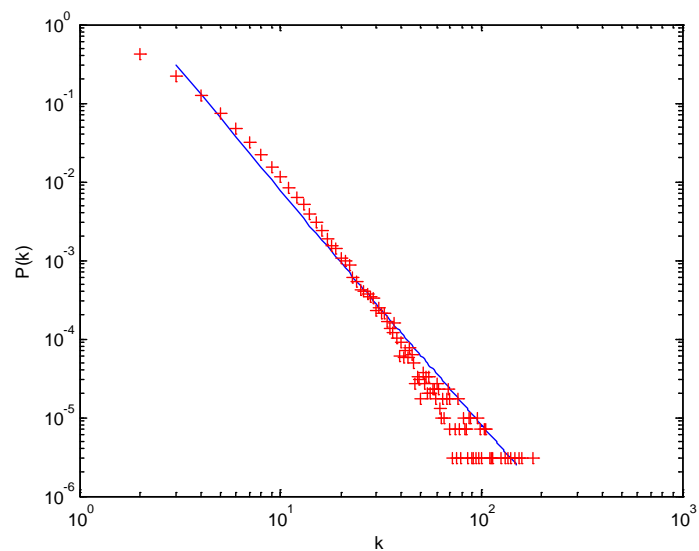

FIG. 5. The hypernetwork model ($\eta_{N(t)}$ is random selected from 1~11) simulation. + denotes the simulation result, the line denotes theoretical prediction.

## V. CONCLUSIONS

This paper proposes the hypernetwork model with the brand effect and competitiveness, which grows with non-uniform. By studying the real-life network, we find out that the probability of obtaining new hyperedges depends not only on the brand effect, but also on its competitiveness and fitness. When a new node with better fitness and more attraction comes into the system, it may be more likely to obtain hyperedges than the old ones. Therefore, besides hyperdegrees, the fitness and competitiveness of the node are two important factors driving the growth of hypernetworks. The preferential mechanism can better reflect the evolving process in a competitive environment and the essence that nodes compete with each other to obtain hyperedges. Through theoretical analysis, we obtain an explicit analytical expression of hyperdegree distributions of the hypernetworks. The theoretical predictions are confirmed by numerical simulations. By setting appropriate parameters, the model can degenerate to competitiveness model, fitness model in complex networks and scale-free hypernetwork model, which demonstrates the universality of the model. The purpose of this paper is to provide a method and some significant reference for the evolving hypernetworks.

Currently the researches on the topological characteristics and evolving mechanism of hypernetworks are just started. Although some scholars have studied the topological properties, the literature on empirical studies of hypernetworks is still rare. Besides, the existing studies simplified the relations and characteristics of nodes and hyperedges. There are still some issues that need to be studied in the future, such as the weighted hypernetwork, directed hypernetwork, aging hypernetwork and hypernetwork combined with exit mechanism.

## ACKNOWLEDGMENTS

The authors acknowledge support from the Shanghai First-class Academic Discipline Project, China (Grant No. S1201YLXK), and supported by the Hujiang Foundation of China (Grant No. A14006).